\RequirePackage{ifpdf}
\documentclass{PoS}

\newcommand{\hi}{H\,{\sc i}}
\def\HIrefer{\par\noindent\hangindent 1.0 truecm\parskip 1pt}

\title{\hi\ science with the Square Kilometre Array}

\ShortTitle{\hi\ Science with the SKA}

\author{Lister Staveley-Smith\\
        (1) International Centre for Radio Astronomy Research (ICRAR), University of Western Australia, 35 Stirling Hwy, Crawley, WA 6009, Australia\\
        (2) ARC Centre for All-sky Astrophysics (CAASTRO)\\
        E-mail: \email{Lister.Staveley-Smith@icrar.org}}

\author{Tom Oosterloo\\
        (1) Netherlands Institute for Radio Astronomy (ASTRON), PO Box 2, 7900 AA Dwingeloo, The Netherlands\\
        (2) Kapteyn Astronomical Institute, Groningen University\\
        E-mail: \email{oosterloo@astron.nl}}

\abstract{The Square Kilometre Array (SKA) will be a formidable instrument for the detailed study of neutral hydrogen (\hi) in external galaxies and in our own Galaxy and Local Group. The sensitivity of the SKA, its wide receiver bands, and the relative freedom from radio frequency interference at the SKA sites will allow the imaging of substantial number of high-redshift galaxies in \hi\ for the first time. It will also allow imaging of galaxies throughout the Local Volume at resolutions of $<100$ pc and detailed investigations of galaxy disks and the transition between disks, halos and the intergalactic medium (IGM) in the Milky Way and external galaxies. Together with deep optical and millimetre/sub-mm imaging, this will have a profound effect on our understanding of the formation, growth and subsequent evolution of galaxies in different environments. This paper provides an introductory text to a series of nine science papers describing the impact of the SKA in the field of \hi\ and galaxy evolution. We propose a nested set of surveys with phase 1 of the SKA which will help tackle much of the exciting science described. Longer commensal surveys are discussed, including an ultra-deep survey which should permit the detection of galaxies at $z=2$, when the Universe was a quarter of its current age. The full SKA will allow more detailed imaging of even more distant galaxies, and allow cosmological and evolutionary parameters to be measured with exquisite precision.}

\FullConference{
Advancing Astrophysics with the Square Kilometre Array\\
June 8-13, 2014\\
Giardini Naxos, Italy}

\begin{document}

\makeatletter
\setbox\@firstaubox\hbox{\small Lister Staveley-Smith}
\makeatother

\section{Introduction} 

The 1944 prediction of the existence of the 21-cm line of atomic hydrogen (\hi; van de Hulst 1945), and its later detection from our Galaxy and other galaxies (Ewen \& Purcell 1951; Muller \& Oort 1951; Kerr, Hindman \& Robinson 1954), was driven by the realisation of the huge scientific potential of observations of this spectral line. From the start, it was clear that atomic hydrogen is not only a fundamental constituent of galaxies whose role has to be studied and understood, but that observations of \hi\ in galaxies can also be used as an important tool for understanding a wide range of phenomena, ranging from large-scale dynamics to the physics of the interstellar medium (ISM). The large amount of \hi-related work over the years following has indeed shown that it is almost impossible to understand galaxies, their structure, the distribution of dark matter and their environmental interactions, without knowing about their \hi\ properties. As a result, improving the observational capabilities in radio astronomy in order to extend the scope of 21-cm studies of galaxies in sensitivity, survey speed, resolution and redshift has always been one of the main science drivers for SKA.  In the following chapters, the main issues the community hopes to address with future \hi\ observations with SKA phase 1 (SKA1) and SKA are outlined. In these chapters, it will be clear that there are strong underlying connections and therefore that there are a number of important themes  that can be addressed by a concerted approach involving observations of our own Galaxy (and its very nearest companions),  detailed studies of nearby galaxies and statistical studies using surveys of a large number of galaxies, out to the highest redshifts available to SKA1 and SKA.

A key goal is to understand the role of \hi\ in the life-cycle of galaxies and how this role evolved over time. How do galaxies acquire gas? What does the interface between galaxies and the intergalactic medium (IGM) look like? How is star formation controlled by gas accretion and, in turn, how does feedback from star formation affect the ISM? Similarly, how is the activity in active galactic nuclei (AGN) connected to \hi\ and how do AGN affect the gas content of galaxies? What is the role of the environment, and of galaxy interactions? With SKA1 and SKA, important progress can be made on answering all these questions, and in particular, how the balance between all the different phenomena has evolved over time.

With current instruments, \hi\ observations of very gas-rich galaxies are possible to redshifts $z\sim 0.2$, or look-back times up to around 2.4 Gyr, but only for a few objects and using long integration times (Zwaan, van Dokkum \& Verheijen 2001; Catinella et al. 2008; Verheijen et al. 2007; Fern\'{a}ndez et al. 2013). With the SKA Pathfinders, huge samples ($>10^5$ galaxies) will be available locally, and significant samples ($>10^4$ galaxies) will be detected to look-back times of $\sim 5$ Gyr. With the improved sensitivity and survey speed offered by SKA1 and SKA, the possibilities to detect and image larger samples of galaxies to even larger look-back times ($7-11$ Gyr) are dramatically improved. Importantly, the high sensitivity of SKA will allow the spatial resolution of the distribution and kinematics of the \hi, as well as the measurement of global quantities, such \hi\ mass and linewidth. This will open up many new ways of investigating the role of \hi\ in galaxy evolution. We must not forget how important it will be to combine these \hi\ data with large galaxy surveys in other wavebands to obtain a complete picture of galaxy evolution. Conversely, the interpretation of those other data sets will not be complete without the matching information available from the SKA \hi\ surveys.

\section{The life-cycle of galaxies}

At a more detailed level, it is necessary to understand the gas cycle in galaxies: how galaxies acquire their gas to fuel star formation and how star formation affects the state of gas in galaxies and mediates the accretion of gas onto galaxies. It is important to understand  the interplay  between the various phases of the gas (ionised, atomic, molecular) and how this connects to star formation. To make progress with modelling galaxy evolution, the accurate  physics of the ISM on the smallest scales has to be understood. With the widefield capability of the SKA (an individual beam is $\sim1$ deg$^2$ at 1.4 GHz), this can be studied in the Galaxy and in the Magellanic Clouds in unprecedented detail. Detailed observations of nearby galaxies will also be required. With the sensitivity of the SKA1, finally we will be able to study the structure and the kinematics of the \hi\ in nearby galaxies, and its relation to star formation and other constituents of the ISM, at spatial resolutions similar to that possible in other wavebands. This will be a major breakthrough. The synergy with other new facilities in the southern hemisphere, such as the Atacama Large Millimeter/sub-millimeter Array (ALMA), the Large Synoptic Survey Telescope (LSST), the multi-field IFU on the Anglo-Australian Telescope (HECTOR), the upcoming 30-m class optical/infrared telescopes, as well as space observatories such as Euclid and eROSITA will lead to significant advances in the field.

An exciting prospect for future SKA observations is having sufficient sensitivity and spatial resolution to detect, for the first time, how galaxies are embedded in the diffuse, low column-density IGM. All current models of galaxy evolution predict that galaxies are surrounded by large gaseous halos from which the galaxies accrete gas to feed their star formation. At column densities below roughly $10^{19}$ cm$^{-2}$, \hi\ is not self-shielding against the intergalactic radiation field. Around this limit,  when going from high to low column density, the ionisation balance quickly changes and the IGM goes from mainly neutral to mainly ionised and \hi\ column densities quickly drop. Therefore, to observe these gaseous halos, we have to be able to detect \hi\ column densities of $10^{18}$ cm$^{-2}$ or lower at a spatial resolution matching  the structure of the filaments. This has been a major barrier for studying the IGM near galaxies with current instruments. SKA1 with its dense, inner array configuration, has the potential to detect and image the gaseous interface between galaxies providing a major step forward in understanding galaxy evolution.

Finally, the low-RFI environment at the SKA sites provides unique observational capabilities to detect \hi\ in emission at moderate redshifts ($z \sim 1$) in phase 1. However, for absorption studies, detections at higher redshifts will readily be possible with SKA1-MID for $z<3$ and SKA1-LOW for $z>3$. Absorption due to intervening \hi\ clouds  along the line-of-sight, as well as \hi\ absorption associated with the ISM surrounding an AGN can be studied in this way. Studies of the intervening \hi\ absorption will allow a blind study of the evolution in the properties of the ISM in galaxies from high redshift to the present. Studies of the associated \hi\ absorption will show how feedback effects of AGN activity on the ISM of galaxies change with redshift. In local galaxies, this activity is often observed in the form of fast, AGN-driven outflows of large amounts of cold gas. Such feedback effects are thought to be important for regulating star formation as well as for the  growth of super massive black holes. 

\section{The \hi\ Science Case}

\begin{figure}[ht]
\begin{center}
\includegraphics[angle=270,width=\textwidth]{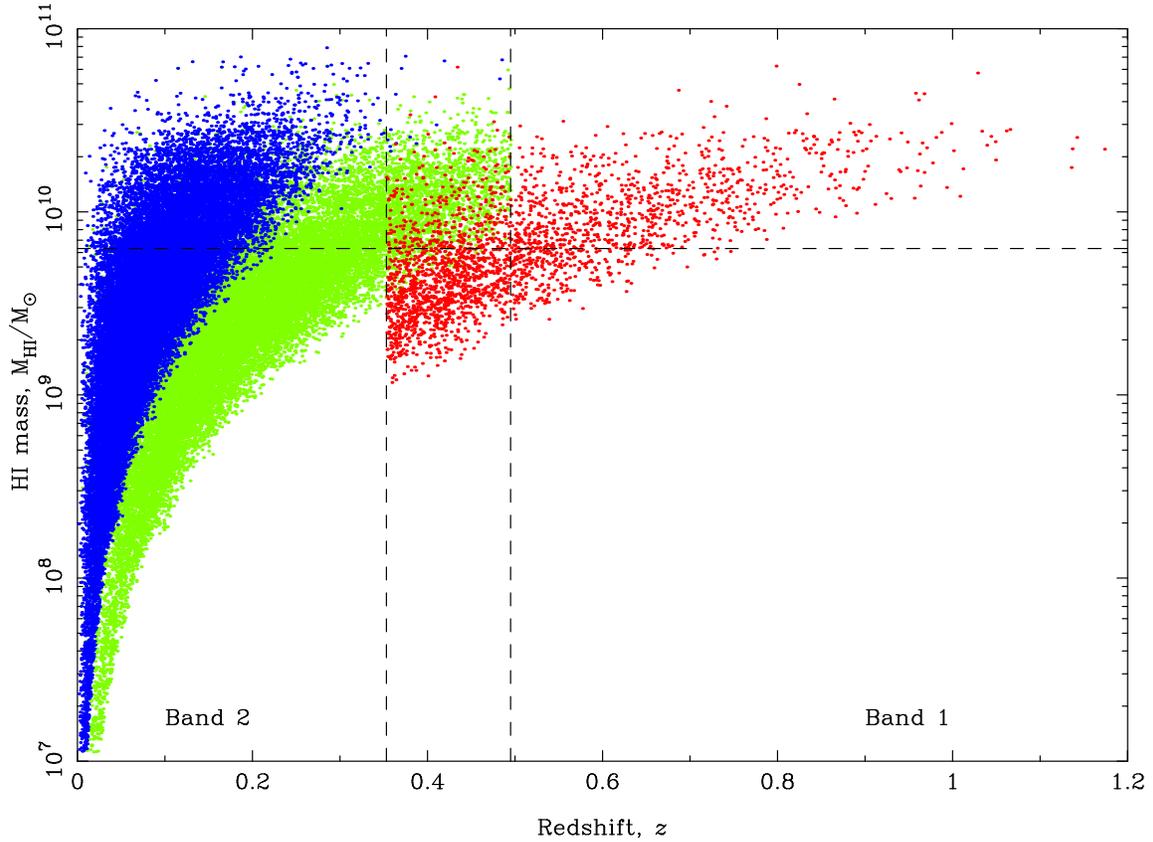}
\end{center}
\caption{A simulation which shows the distribution with redshift of \hi\ masses of galaxies likely to be detected in the representative tiered set of 1,000 hr surveys discussed in the text and in the following table. The `medium wide' (blue), `medium deep' (green) and `deep' (red) surveys are plotted. The vertical dashed lines refer to the (overlapping) band edges, as specified in Dewdney et al. (2013). The horizontal line represents the value of $M_{HI}^*$ at $z=0$. A non-evolving 2-dimensional stepwise maximum likelihood (2DSWML) \hi\ mass and velocity width function  (Zwaan et al. 2005) is assumed for the purpose of this simulation.}
\label{hioverview:survey.fig}
\end{figure}

The following chapters concentrate on the far-reaching impact of SKA1 on our understanding of the formation and evolution of galaxies (Blyth et al. 2015; Power et al. 2015; Meyer et al. 2015) including the build-up of angular momentum (Obreschkow et al. 2015), the accretion of gas into the outskirts of galaxies (Popping et al. 2015), the detailed understanding of the ISM of galaxies (de Blok et al. 2015), and the impact of AGN on the large-scale gas distribution in galaxies (Morganti et al. 2015). The impact of SKA1 on our understanding of the local Galactic ISM and halo is also explored, including detailed mapping via absorption-line studies of the density and temperature of the cold neutral medium (McClure-Griffiths et al. 2015; Oonk et al. 2015). This impressive set of science cases and chapters has been facilitated by the \hi\ and Galaxy Evolution working group of the SKA Science Working Group. Strong links are noted with the Cosmology working group (Maartens et al. 2015) who have explored use of the SKA in measuring cosmological parameters and exploring evolution through galaxy surveys and intensity mapping (Bull et al. 2015; Santos et al. 2015).

\section{Possible SKA1 \hi\ surveys}

As with the SKA pathfinders, a major part of the science output of the SKA is likely to come from the allocation of significant amounts of telescope time (1,000 hrs, or greater) to Key Science Projects. In order to maximise the \hi\ science output of the SKA, it is desirable for these to be complementary in nature between science cases. For example, a natural suite of surveys may include: (1) a wide area survey for studies of cosmology; (2) medium deep surveys for studies of environmental dependence and individual galaxies; and (3) a deep survey for studies of evolution. \hi\ chapter authors have indicated their range of preferences based on the fiducial baseline design of Dewdney et al. (2013). The recent re-baseline only has a modest effect (less than a factor of two in observing time) on science goals best achieved with SKA1-MID, but a more dramatic effect on goals that require SKA1-survey. Nevertheless, on the basis that some SKA1-survey science can still be achieved with SKA1-MID, we retain the moniker as used in the individual science chapters.

\begin{table}[b!]
\begin{center}
\small
\begin{tabular}{lccccccl}

\hline
Survey & $\Omega$ & Freq-    &  Resol-    & $N$ &  < $z$ > & $N_{\rm HI}$ & Science chapter \\
       &          & ency$^1$ &  ution$^2$ &     &  ($z_{lim}$)  &  $10^{20}$        &  \\
       & deg$^2$  &  MHz     &            &     &           & cm$^{-2}$ &            \\ \hline
        & \\
Galaxy/MS  & 400& 1418-1422  &     $5''$   & 4,000  &  & & McClure-Griffiths et al. \\
(absorption)            &     &           &            &   &   & & (2015)\\ 
Extragalactic            & 1000& 350-1050  &     $2''$   & 5,000  & 1(3) & & Morganti et al. (2015) \\
(absorption)     & 1000& 200-350$^3$ &    $10''$ &  ? & 4(6) &  & ~~~~~~~'' \\ 
            &     &           &            &   &   & & \\  \hline
 & \\
Galaxy/MS   & 600& 1418-1422 &  $10''$-$1'$  &   &  & 2 & McClure-Griffiths et al. \\
            &     &           &            &   &   & & (2015); Oonk et al. (2015)\\ 
Medium wide & 400 & 950-1420  &    $10''$   & 34,000 & 0.1 & 2 & Meyer et al. (2015); \\
            &     &           &             &   &  (0.3) & & Obreschkow et al. (2015);\\
            &     &           &             &   &   & & Popping et al. (2015)\\
Medium deep & 20  & 950-1420  &    $5''$    & 25,000  & 0.2 & 0.6 & Meyer et al. (2015); \\
            &     &           &             &   &  (0.5) & & Obreschkow et al. (2015);\\
            &     &           &             &   &   & & Popping et al. (2015)\\
Deep        & 1   & 600-1050  &     $2''$   &  2,600 & 0.5 & 0.4 & Blyth et al. (2015); \\
            &     &           &             &   & (1)  & & Meyer et al. (2015); \\
            &     &           &             &   &   & & Obreschkow et al. (2015);\\
            &     &           &             &   &   & & Power et al. (2015)\\
Targeted & -- & 1400-1420 &   $3''$-$1'$  & 50 & 0.002 & 0.5 & de Blok et al. (2015); \\
            &     &           &             &   & (0.01)  & & Popping et al. (2015) \\ \hline
\end{tabular}
\end{center} 
{\scriptsize
$^1$SKA1-MID band definition is 350--1050 MHz (band 1) and 950--1760 MHz (band 2); SKA1-LOW band definition is 50--350 MHz.\\
$^2$Typical angular resolution; value changes with frequency and (for \hi\ emission) column-density sensitivity threshold.\\ 
$^3$SKA1-LOW.
}
\caption{Fiducial 1,000-hr SKA1 \hi\ surveys and the science cases that they would facilitate. The list is not complete, but illustrates the tiered structure necessary to allow a comprehensive, multi-faceted study of galaxies and their evolution in a range of environments and over a range of redshifts. The SKA1 telescope parameters are those resulting from the 2015 re-baseline (see text for further details). For non-targeted emission-line surveys, the predicted number of galaxies $N$ (column 5) assumes an invariant 2DSWML \hi\ mass and velocity width function (Zwaan et al. 2005) and assumes a detection at 5-$\sigma$ significance or higher, at $10''$ resolution. The mean sample redshift <$z$> and redshift upper limit $z_{lim}$ are indicated (column 6). The corresponding 5-$\sigma$ column density sensitivity $N_{\rm HI}$ for extended objects at <$z$> is also indicated (column 7), assuming a rest-frame velocity width of 20 km s$^{-1}$. Angular resolutions finer than $10''$ are available down to the approximate value indicated in column 4, but only for high column densities. The prediction follows methodology similar to that of Duffy et al. (2012). }
\label{hioverview:survey.tab}
\end{table}

In suggesting the nested set of surveys in Table~\ref{hioverview:survey.tab} and Figure~\ref{hioverview:survey.fig}, we have referred to the individual chapters for guidance. We have used the simulations of Popping et al. (2015) as a guide to the SKA1 sensitivity as a function of resolution, taking into account a `re-baselining' scale factor of 1.3 and, for SKA1-MID band 1 only, a further band-average sensitivity loss of $\sim1.7$ due to system temperature and antenna efficiency (Dewdney et al. 2013). We have accounted for the change of primary beam with frequency. In selecting surveys, we also need to be cognisant of the capabilities of the various SKA precursors and pathfinders, ASKAP, FAST, JVLA, MeerKAT and WSRT/APERTIF, some of which have sensitivities or survey speed not very different from SKA1 at lower angular resolutions. In particular, an all-sky emission-line survey with SKA1 will not be better than those currently planned for the pathfinders unless more integration time can be devoted than indicated in Table~\ref{hioverview:survey.tab}.

\section{Commensal SKA1 Observations}

\begin{table}[h]
\begin{center}
\small
\begin{tabular}{lccccccl}

\hline
Survey & $\Omega$ & Freq- &  Resol- & $N$ & < $z$ > & $N_{\rm HI}$& Science chapter \\
       &          & ency  &  ution  &     &  ($z_{lim}$)         & $10^{20}$ & \\
       & deg$^2$  &  MHz      &     &     &          & cm$^{-2}$     &                 \\ \hline
        & \\
All-sky     & 20,000 & 950-1420  &  $15''$  &  550,000 &  0.06 & 2 & McClure-Griffiths et al. (2015); \\
            &        &           &          &   & (0.3)  & & Obreschkow et al. (2015);\\ 
            &        &           &          &   &   & & Bull et al. (2015)\\ 
Wide        & 5,000   & 950-1420  &  $10''$ &  340,000 & 0.1 & 2 & Meyer et al. (2015); \\
            &        &           &          &   & (0.5)  & & Obreschkow et al. (2015);\\
            &        &           &          &   &   & & Popping et al. (2015)\\
Ultra deep  & 1      & 450-1050  &  $2''$   &  23,000 & 0.7 & 0.2 & Blyth et al. (2015); \\
            &     &           &             &         & (2)  &                 & Meyer et al. (2015); \\
            &     &           &             &         &   &                 & Obreshkow et al. (2015);\\
            &     &           &             &         &   &                 & Power et al. (2015)\\ \hline
\end{tabular}
\end{center} 
\caption{Commensal 10,000 hr SKA1 surveys across science areas allow much greater observing times for a smaller number of surveys. Here we list some possible commensal surveys, each with a total integration time of 10,000 hrs. The scientific goals of such surveys extend well beyond the \hi\ and cosmology science chapters listed here. The numbers of detectable galaxies and 5-$\sigma$ column density limits at the mean redshift <$z$> are estimated at $15''$ resolution for the all-sky survey, and $10''$ otherwise. Column density limits at $1'$ resolution are $\sim 33$ times better (Popping et al. 2015).}
\label{hioverview:commensal.tab}
\end{table}

With SKA1, greater sensitivities are possible for surveys that are commensal across science themes. Whereas Table~\ref{hioverview:survey.tab} indicates that dedicated survey times of $\sim 1,000$ hrs will meet key \hi\ science goals such as investigating the cosmic evolution of galaxies at high angular resolution to $z\sim1$, a commensal survey could in principle have a factor of $\sim 10$ more integration time. This would permit a large-area \hi\ survey, which cannot otherwise be undertaken with SKA1 in 1,000 hrs because of column density sensitivity limits. This would enable `SKA1-survey' outcomes discussed by various chapter authors (Obreschkow et al. 2015; McClure-Griffiths et al. 2015; Meyer et al. 2015), as well as \hi\ cosmology outcomes (Bull et al. 2015), and important commensal pulsar, continuum and magnetism science outcomes (e.g. Keane et al. 2015; Norris et al. 2014; Gaensler et al. 2015). Another possibility is an ultra-deep survey which would allow detection of \hi\ emission from galaxies at $z\sim 2$ (Blyth et al. 2015) as well as allowing studies of continuum star formation rates in galaxies at the highest redshifts (Jarvis et al. 2014). Similarly, deep and highly resolved \hi\ observations of nearby galaxies would also yield deep continuum and polarisation data. These could be used to further explore the physics of the continuum-star formation rate relation, rotation measures, and magnetic field configuration. Such observations would also have very deep column density limits ($<10^{18}$ cm$^{-2}$) at $1'$ resolution. We list a subset of three possible \hi\ commensal surveys in Table~\ref{hioverview:commensal.tab}. Although only one or two such commensal surveys are likely to be possible, they would be a significant step towards the full SKA.

\section{The full SKA}

One of the early headline science goals of the SKA was to image \hi\ in a `Milky Way-type' galaxy at a cosmologically significant ($z\sim 2$) redshift (Taylor \& Braun 1999). In the updated science case edited by Carilli \& Rawlings (2004), the possibility of retaining such sensitivity was emphasised (van der Hulst et al. 2004). However, it was also pointed out that with the large field-of-view of new-technology radio telescopes, it is possible to observe extremely large samples of galaxies at lower resolution to undertake 21cm cosmology. The so-called `billion galaxy survey' (Abdalla \& Rawlings 2005) was proposed to investigate the nature of dark energy over the crucial redshift range $1.5 > z > 0$. Although this remains a clear science goal for the full SKA, it receives less attention in this volume which focusses on SKA1. Furthermore, this needs re-evaluation in light of projects such as Euclid and DES which have been funded in the meantime, and the potential of other novel techniques that can be applied to SKA data such as Intensity Mapping (Chang et al. 2010). The full SKA would also transform our understanding of the galaxy evolution and dynamics over the range of redshifts even beyond that currently accessible to the largest optical telescopes.

For example, an ultra-deep single-pointing with an SKA of 10 times the sensitivity of SKA1 would be able to detect and resolve galaxies, or compact groups of galaxies, with \hi\ masses in excess of $10^{10}$ M$_{\odot}$ out to the $z=3$ redshift limit of band 1 in 1000 hrs, when the Universe was a sixth of its current age. Similarly, a hemispheric survey using an SKA with 100 times the survey speed of SKA1 would be able to detect around $2.5\times10^7$ galaxies in \hi\ out to the $z=0.5$ redshift limit of band 2 in 10,000 hrs, and almost a million galaxies in the redshift range $0.35 < z < 0.7$ with similar band 1 integration times. The more ambitious `dark energy' evolution survey (Abdalla, Blake \& Rawlings 2010) of a similar duration would require a telescope whose survey speed is over 4,000 times greater than the band 1 survey speed of SKA1, and would require large numbers of antennas, phased array feeds or aperture arrays. If feasible, such a survey could detect $10^8$ galaxies with redshifts $z>0.35$, of which over $2\times 10^6$ would be at $z>1$ in a no-evolution scenario.

\section{Acknowledgments}

It is a pleasure to thank the members of the SKA \hi\ and Galaxy Evolution working group who have contributed to this overview and for writing the chapters contained within this science book.

\section*{References}


\HIrefer{Abdalla, F. B., Blake, C., Rawlings, S.,  2010,  MNRAS,  401,  743.}

\HIrefer{Abdalla, F. B., Rawlings, S.,  2005,  MNRAS,  360,  27.}

\HIrefer{Blyth, S.-L., van der Hulst, J. M., Verheijen, M. A. W., et al.,  2015, "Exploring Neutral Hydrogen and Galaxy Evolution with the SKA", in proc. {\em Advancing Astrophysics with the Square Kilometre Array}, PoS(AASKA14)128}

\HIrefer{Bull, P., Camera, S., Raccanelli, A., et al.,  2015,  "Measuring baryon acoustic oscillations with future SKA surveys",  in proc. {\em Advancing Astrophysics with the Square Kilometre Array}, PoS(AASKA14)024}

\HIrefer{Carilli C.L., Rawlings S. 2004, eds. "Science with the Square
Kilometre Array", NewAR, 48, 979.}

\HIrefer{Catinella, B., Haynes, M. P., Giovanelli, R., Gardner, J. P., Connolly, A. J.,  2008,  ApJ,  685,  L13.}

\HIrefer{Chang, T.-C., Pen, U.-L., Bandura, K., Peterson, J. B.,  2010,  Nature,  466,  463.}

\HIrefer{de Blok, W. J. G., Fraternali, F., Heald, G. H., et al.,  2015, "The SKA view of the Neutral Interstellar Medium in Galaxies", in proc. {\em Advancing Astrophysics with the Square Kilometre Array}, PoS(AASKA14)129}

\HIrefer{Dewdney, P. E., Turner, W., Millenaar, R., McCool, R., Lazio, J., Cornwell, T. J., 2013, SKA1 System Baseline Design (SKA Office), SKA-TEL-SKO-DD-001 revision 1.}

\HIrefer{Duffy, A. R., Meyer, M. J., Staveley-Smith, L., et al.,  2012,  MNRAS,  426,  3385.}

\HIrefer{Ewen, H. I., Purcell, E. M.,  1951,  Nature,  168,  356.}

\HIrefer{Fern\'{a}ndez, X., van Gorkom, J. H., Hess, K. M., et al.,  2013,  ApJ,  770,  L29.}

\HIrefer{Gaensler, B. M., Agudo, I., Akahori, T., et al.,  2015,  "Broadband Polarimetry with the Square Kilometre Array: A Unique Astrophysical Probe", in proc. {\em Advancing Astrophysics with the Square Kilometre Array}, PoS(AASKA14)103}

\HIrefer{Jarvis, M. J., Seymour, N., Afonso, J., et al.,  2015, "The star-formation history of the Universe with the SKA", in proc. {\em Advancing Astrophysics with the Square Kilometre Array}, PoS(AASKA14)068}

\HIrefer{Keane, E. F., Bhattacharyya, B., Kramer, M., et al.,  2015,  "A Cosmic Census of Radio Pulsars with the SKA", in proc. {\em Advancing Astrophysics with the Square Kilometre Array}, PoS(AASKA14)040}

\HIrefer{Kerr, F. J., Hindman, J. F., Robinson, B. J.,  1954,  AuJPh,  7,  297.}

\HIrefer{Maartens, R., Abdalla, F. B., Jarvis, M., et al.,  2015,  "Overview of Cosmology with the SKA", in proc. {\em Advancing Astrophysics with the Square Kilometre Array}, PoS(AASKA14)016}

\HIrefer{McClure-Griffiths, N. M., Stanimirovic, S., Murray, C. E., et al., 2015, "Galactic and Magellanic Evolution with the SKA", in proc. {\em Advancing Astrophysics with the Square Kilometre Array}, PoS(AASKA14)130}

\HIrefer{Meyer, M., Robotham, A., Obreschkow, D., Driver, S., Staveley-Smith, L., Zwaan, M.,  2015,  "Connecting the Baryons: Multiwavelength Data for SKA HI Surveys", in proc. {\em Advancing Astrophysics with the Square Kilometre Array}, PoS(AASKA14)131}

\HIrefer{Morganti, R., Sadler, E. M., Curran, S. J., 2015, "Cool Outflows and HI absorbers with SKA", in proc. {\em Advancing Astrophysics with the Square Kilometre Array}, PoS(AASKA14)134}

\HIrefer{Muller, C. A., Oort, J. H.,  1951,  Nature,  168,  357.}

\HIrefer{Norris, R. P., Basu, K., Brown, M., et al., 2015, "The SKA Mid-frequency All-sky Continuum Survey: Discovering the unexpected and transforming radio-astronomy", in proc. {\em Advancing Astrophysics with the Square Kilometre Array}, PoS(AASKA14)086}

\HIrefer{Obreschkow, D., Meyer, M., Popping, A., Power, C., Quinn, P., Staveley-Smith, L.,  2015,  "The SKA as a Doorway to Angular Momentum", in proc. {\em Advancing Astrophysics with the Square Kilometre Array}, PoS(AASKA14)138}

\HIrefer{Oonk, J. B. R., Morabito, L. K., Salgado, F., et al.,  2015,  "The Physics of the Cold Neutral Medium: Low-frequency Radio Recombination Lines with the Square Kilometre Array", in proc. {\em Advancing Astrophysics with the Square Kilometre Array}, PoS(AASKA14)139}

\HIrefer{Popping, A., Meyer, M., Staveley-Smith, L., Obreschkow, D., Jozsa, G. I., Pisano, D. J.,  2015,  "Observations of the Intergalactic Medium and the Cosmic Web in the SKA era", in proc. {\em Advancing Astrophysics with the Square Kilometre Array}, PoS(AASKA14)132}

\HIrefer{Power, C., Lagos, C. D. P., Qin, B., et al.,  2015,  "Galaxy Formation \& Dark Matter Modelling in the Era of the Square Kilometre Array", in proc. {\em Advancing Astrophysics with the Square Kilometre Array}, PoS(AASKA14)133}

\HIrefer{Santos, M. G., Bull, P., Alonso, D., et al.,  2015,  "Cosmology from a SKA HI intensity mapping survey", in proc. {\em Advancing Astrophysics with the Square Kilometre Array}, PoS(AASKA14)019}

\HIrefer{Taylor, A. R., Braun, R.,  1999, eds. `Science with the Square Kilometer Array: a next generation world radio observatory'.}

\HIrefer{van de Hulst, H. C., 1945, Ned.Tijd.Natuurkunde, 11, 21.}

\HIrefer{van der Hulst J.M., Sadler E.M., Jackson C.A., Hunt L.K., Verheijen M., van Gorkom J.H. 2004, NewAR, 48, 1221.}

\HIrefer{Verheijen, M., van Gorkom, J. H., Szomoru, A., Dwarakanath, K. S., Poggianti, B. M., Schiminovich, D.,  2007,  ApJ,  668,  L9.}

\HIrefer{Zwaan M. A., Meyer M. J., Staveley-Smith L., Webster R. L., 2005, MNRAS, 359, 30}

\HIrefer{Zwaan, M. A., van Dokkum, P. G., Verheijen, M. A. W.,  2001,  Sci,  293,  1800.}

\end{document}